\pdfoutput=1
\documentclass[structabstract]{aa}  
\usepackage{txfonts}
\usepackage{graphicx}
\begin{document}
\newcommand{\chg}{}
\newcommand{\hhh}{\mbox{H$_2$ }}
\newcommand{\ma}{\mbox{m\AA}}
\newcommand{\ms}{\mbox{m s$^{-1}$}}
\newcommand{\kms}{\mbox{km s$^{-1}$}}
\title{QSO 0347-383 and the invariance of m$_p$/m$_e$ in the course of cosmic
time \thanks{Based on observations taken at  ESO   Paranal Observatory, Program  083.A.0733}}

\author{
M. \,Wendt \inst{1} \and P. \, Molaro\inst{2}
          }

\institute{Institut f\"ur Physik und Astronomie
, Universit\"at Potsdam, 14476 Golm, Germany\\
\email{mwendt@astro.physik.uni-potsdam.de}
\and
Osservatorio Astronomico di Trieste, Via G.\,B.\,Tiepolo 11,
34131 Trieste, Italy
}

\authorrunning{M.~Wendt \and P.~Paolo}

\date{Received \today; accepted -- --}
\abstract
{The variation of the dimensionless fundamental physical constant  $\mu=m_p/m_e$ -- the proton to electron mass ratio --
 can be constrained via observation of Lyman and Werner lines of molecular hydrogen in
the spectra of  damped Lyman alpha systems (DLAs) in the line of sight to distant QSOs. }
{Our intention is to max out the possible precision of quasar absorption spectroscopy with regard to the investigation of the variation of the proton-to-electron mass-ratio $\mu$. The demand for precision requires an  understanding of the errors involved  and effective techniques to handle present
systematic errors.}
{
An analysis based on UVES high resolution 
data sets of QSO 0347-383 and its DLA  is put forward and
new approaches to some of the steps involved in the data analysis are introduced. We apply corrections for the observed offsets between discrete spectra and for the first time we find indications for inter-order distortions.}
{Drawing on VLT-UVES observations of QSO 0347-383 in 2009 our  analysis
yields $\Delta\mu/\mu = (4.3 \pm 7.2) \times
10^{-6}$ at $z_{\mathrm{abs}}=3.025$. 
}
{Current analyzes tend to underestimate the impact of systematic errors.  Based on the scatter of the measured redshifts and the corresponding low significance
of the redshift-sensitivity correlation we estimate the limit of accuracy of
line position measurements to $\sim$ 220 \ms, consisting of roughly 150 \ms due to
the uncertainty of the absorption line fit and about 150 \ms allocated to 
systematics related to instrumentation and calibration.}


\maketitle

\section{Introduction}
The Standard Model of particle physics  contains    several 
fundamental constants whose values
cannot be predicted by theory and   need to  be measured through 
experiments (Fritzsch \cite{Fritzsch09}).     They are the mass of the elementary particles and the  dimensionless coupling constants which are assumed time-invariant  although   in  theoretical models    which seek  to unify
the four forces of nature they  vary naturally on cosmological
scales. 
 The fine-structure constant $\alpha \equiv e^2/(4\pi \epsilon_0 \hbar c)$ and 
the proton-to-electron mass ratio, $\mu  = m_p / m_e$  are two constants that can be probed in the laboratory as well as  in the  Universe by means of 
observations of  absorption lines due to intervening systems in the spectra of distant QSOs and have 
 been the
subject of numerous studies. The former is related to the electromagnetic force while the latter  is  sensitive primarily   to
the quantum chromodynamic scale (see, i.e., Flambaum \cite{Flambaum04}). 

A probe of the variation of $\mu$ could be obtained by comparing
relative frequencies of the electro-vibro-rotational lines of \hhh
as first applied by Varshalovich and Levshakov (\cite{Varshalovich93}) after Thompson (\cite{Thompson75})
 proposed the general approach to utilize molecule transitions for $\mu$-determination.
The original paper by Thompson (\cite{Thompson75}) did not take into account the different
sensitivities within the molecular bands, which is the key of the modern approach. 

The method is based on the fact that the wavelengths of vibro-rotational
lines of molecules depend on the reduced mass, M, of the molecule.
For molecular hydrogen M $= m_p/2$ so that the comparison of an observed vibro-rotational 
spectrum with its present analog will  give information on the variation of $m_N$ and $m_e$.
Comparing electro-vibro-rotational lines with different
sensitivity coefficients gives a measurement of $\mu$.

%
The observed wavelength $\lambda_{\mathrm{obs},i}$  of any given
line in an absorption system at the redshift $z$ differs from the local
rest-frame
wavelength $\lambda_{0,i}$  of the same line in the laboratory according to the
relation 

 \begin{equation}
   \lambda_{\mathrm{obs},i}  =  \lambda_{0,i} (1 + z)(1+ K_i \frac{\Delta \mu}{
\mu}),
 \end{equation}   
where $K_i$ is the sensitivity coefficient of the $i$th component computed
theoretically for the Lyman and Werner bands of the H$_2$
molecule (Varshalovich \& Levshakov \cite{Varshalovich93},
 Varshalovich \& Potekhin \cite{Varshalovich95},
Potekhin et al. \cite{Potekhin98},
Meshkov et al. \cite{Meshkov07},
Ubachs et al. \cite{Ubachs07}).

It is useful to measure  variations in velocities with comparison to the redshift of a given system defined by the redshift position of the lines with $K_i \approx0$, then introducing the reduced redshift  $\zeta_i$:
\begin{equation}
 \frac{\Delta V_i}{c} \approx \zeta_i \equiv \frac{z_i - z}{1+z} = K_i \frac{\Delta\mu}{\mu}. \label{eq_LBLFM}
\end{equation}
 The velocity shifts of  the lines are   linearly proportional to $\Delta\mu/\mu$ which  can be measured  through a regression analysis in the $\Delta V_i - K_i $
plane.

This method was used   to obtain  a bound  on
the secular variation of the electron-to-proton mass ratio at $\Delta \mu / \mu = (-1.8 \pm 3.8) \times 10^{-5}$  from observations of
the newly discovered H$_2$ absorption systems at $z_{abs}$ =3.0 towards QSO 0347-383 
   (Levshakov et al. \cite{Levshakov02}).
 Subsequent   measures  of  the  absorption systems of  QSO 
0347-383  and QSO  1232+082 provided a strong indication of a variation $(2.4 \pm 0.6) \times
10^{-5}$, i.e.
at  3.5 $\sigma$  (Reinhold et al. \cite{Reinhold06}, Ubachs et al. \cite{Ubachs07}). 
Earlier works (Ivanchik et al. \cite{Ivanchik05}) also find hints for variation, but are still dominated by inaccuracies 
of the laboratory wavelengths. 
However,  more recently King et al. (\cite{King08}), Wendt \& Reimers
(\cite{Wendt08}), Thompson et al. (\cite{Thompson09a}a), Wendt and Molaro (\cite{Wendt11}),
King et al. (\cite{King11}) with  $\Delta \mu / \mu = (0.3 \pm 3.7) \times 10^{-6}$ 
at $z_{abs}$ = 2.811 towards PKS 0528-250 and Bagdonaite et al. (\cite{Bagdonaite12}) 
with $\Delta \mu / \mu = (-6.8 \pm 27.8) \times 10^{-6}$ 
at $z_{abs}$ = 2.426 towards QSO 2348-011 reported a result in agreement with no
variation.

The  more stringent limits on   $\Delta \mu / \mu$ have been found    from the combination of three H$_2$ systems at $\Delta\mu / \mu  = (2.6\pm 3.0_{\rm stat})\times10^{-6}$( King et al.
\cite{King08}) and taking into account additional transitions from deuterated molecular hydrogen (HD)
 in King et al. (\cite{King11}).
A fourth system has provided $\Delta\mu / \mu  = (+5.6\pm 5.5_{\rm stat}\pm
2.9_{\rm sys})\times10^{-6}$
(Malec et al. \cite{Malec10}).


 An independent method relies on  the inverse spectrum of ammonia as shown by   Flambaum \& Kozlov
(\cite{Flambaum07}).
Ammonia NH$_3$  inversion transitions  are very
sensitive to changes in $\mu$ due to a tunneling
effect. The sensitivity coefficient of the inversion transition can be 
 almost two orders of magnitude 
more sensitive to $\mu$-variation than H$_2$ molecular rotational frequencies. 
Thus by comparing the  inversion
frequency of NH$_3$(1,1) with a  rotational frequency of another
co-spatial molecule  it is possible to bind  a
variation of $\mu$.

 Flambaum
\& Kozlov (\cite{Flambaum07})
 combine three detected NH$_3$  absorption spectra from
B0218+357 with rotational spectra of CO, HCO$^+$, and HCN to place a limit of
(0.6$\pm$ 1.9) $\times$ 10$^{-6}$  for a look-back time of 6  Gyr (redshift z =
0.68).
 Murphy et
al.
(2008)  with newly obtained high
signal-to-noise rotational spectra of HCO$^+$ and HCN obtained   $<$ 1.8
$\times$
10$^{-6}$ at a 95\% CL. 

Henkel et al. (2009) obtained   a firm upper
limit  of 10$^{-6}$ for a look-back time of 7 Gyr (z=0.89) towards PKS 1830-211. 
This method is limited to low redshifts due to the small number of NH$_3$ sources in general 
and to the large line widths and chemical segregation   of different molecules at higher redshifts.
 For sources in the local Milky Way, however, a very 
strict upper limit of $|\Delta\mu/\mu| < 3 \times 10^{-8}$
was found utilizing both the ammonia method 
(Levshakov et al.  \cite{Levshakov10a}, and Levshakov et al.  \cite{Levshakov10b}), 
and the methanol method (Levshakov et al. \cite{Levshakov11}).

Laboratory
experiments  by comparing  the rates between clocks based on hyperfine
transitions
in atoms with   a different dependence on $\mu$    restrict the
 time-dependence of  $\mu$     at 
the level of {$ (\dot{\mu}/ \mu)_{t_0}  = (1.6 \pm 1.7) \times  10^{-15}$
yr$^{-1}$} (Blatt et al. \cite{Blatt08}).

In the following we will concentrate on the   \hhh system observed towards
QSO 0347-383 to trace the
proton-to-electron mass ration $\mu$
at high redshift  ($z_{\mathrm{abs}}=3.025$). 
   The motivation for re-analysis of QSO 0347-383 is given by the dramatically enhanced
quality of the recent data of this quasar for the purpose of setting constraints on 
$\Delta\mu/\mu$. 
The single velocity component in \hhh absorption renders
 QSO 347-383 a prime target to further investigate the impact of wavelength
 calibration issues.
 Trying to reach a sensitivity of few
parts per million everything becomes important and the special
 requirements of the observations as described in the following section are absolutely mandatory.

\section{Data}
\subsection{Observations}
The recent observations of QSO 0347-383 were performed with
UVES on VLT on the nights of September 20-24 2009.
{\chg The journal of these observations  is given in Table \ref{tab:obs2009}. The DIC 2
setting was used with blue setting and the 437 nm grating.}
 The CCDs were not binned with 
 pixel size  of $0.013-0.015$ \AA,  
or $1.12$ \kms\  at 400 nm   along the dispersion direction. The
observations are comprised of 11 exposures on four
successive nights, of which  10 exposures were of  5400 s and one of 3812 s.
Eight of the spectra with setting 437+760 and
three the 437+860 setting,  providing  a coverage in the blue spectral ranges between 373-500 nm.
QSO 0347-383 has no flux below 370 nm due to the Lyman discontinuity of the
$z_{\rm abs} = 3.023$ absorption system.
\begin{table}[h]
\caption[Journal of the observations (2009 data)]{Journal of the observations
(2009 data). Before and after each
spectrum, a 30 s calibration frame was recorded}
\label{tab:obs2009}      
\centering          
\begin{tabular}{c c c c c r c}     
\hline\hline       

 No. & Date  & Time  & $\lambda$ &Exp[s]& DIMM[arcsec]\\ 
\hline                    
 1 & 2009-09-20 & 05:05:46 & 437 & 5400&1.21(0.29)\\
   2 & 2009-09-20 & 08:28:48 & 437 & 3812&1.73(0.23)\\
   3 & 2009-09-21 & 04:45:51 & 437 & 5400&1.15(0.12)\\
   4 & 2009-09-21 & 06:18:45 & 437 & 5400&1.21(0.18)\\
   5 & 2009-09-21 & 07:59:24 & 437 & 5400&1.08(0.09)\\
   6 & 2009-09-22 & 04:41:37 & 437 & 5400&0.97(0.17)\\
   7 & 2009-09-22 & 06:14:25 & 437 & 5400&1.03(0.13)\\
   8 & 2009-09-22 & 07:59:19 & 437 & 5400&1.00(0.14)\\
   9 & 2009-09-23 & 04:24:05 & 437 & 5400&1.36(0.26)\\
  10 & 2009-09-23 & 05:56:49 & 437 & 5400&1.18(0.23)\\
  11 & 2009-09-23 & 07:29:35 & 437 & 5400&0.95(0.23)\\
\hline                  
\end{tabular}
\end{table}
The slit width was set to 0.7\arcsec\, for all observations providing a
mean resolving power
of $\lambda$/$\Delta \lambda$ $\approx$ $66000$. {\chg Within each  order the resolving power  varies by about 15-20 \% being higher at the starting  wavelength 
of  each order. The average seeing along the exposures as recorded by the DIMM at Paranal is given in the last column of  Table \ref{tab:obs2009}. We note, however, that the actual seeing at the UT2 of VLT was significantly better than that recorded by the DIMM.}

\begin{figure}[h]
\resizebox{\hsize}{!}{\includegraphics[clip=true]{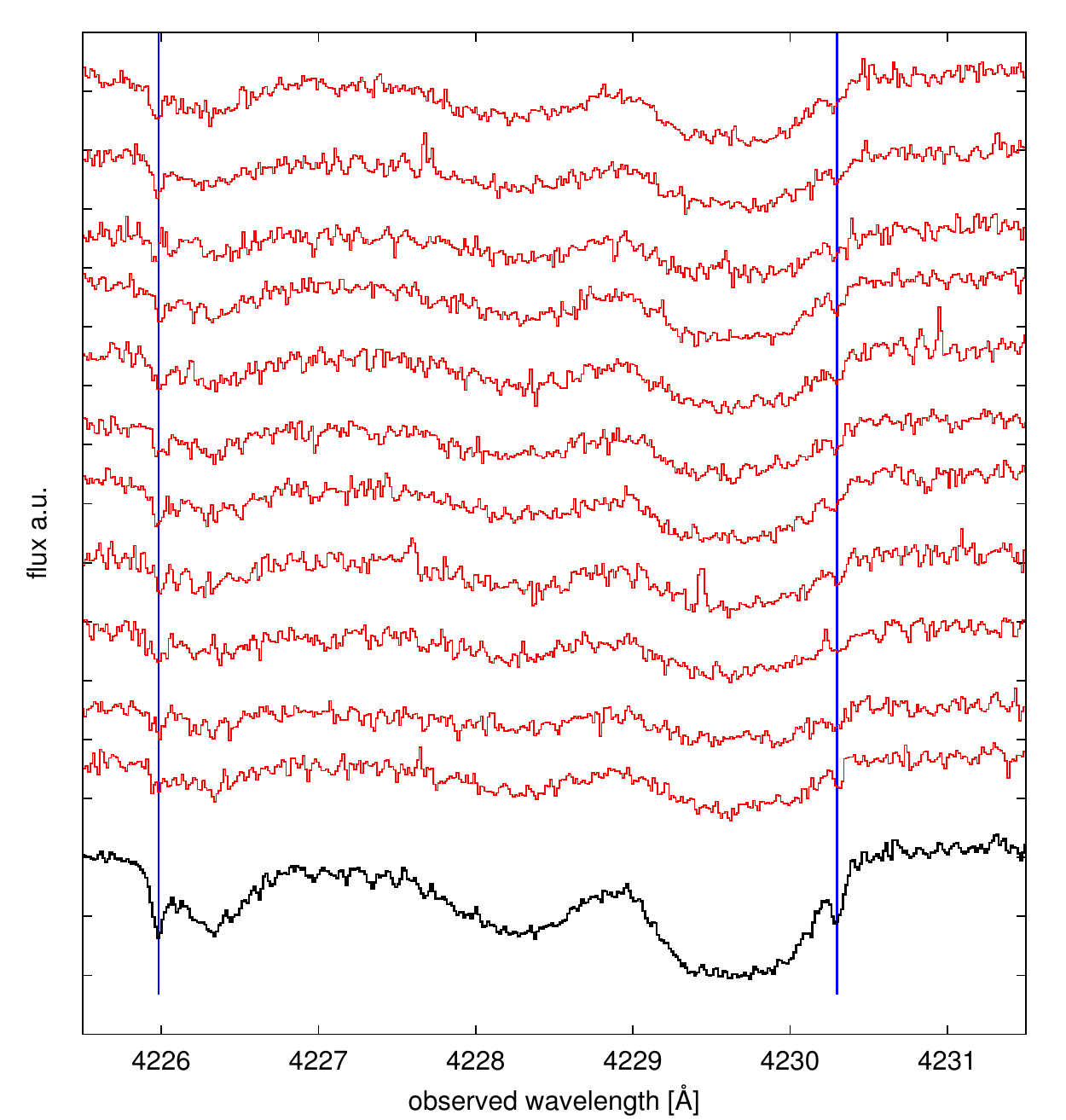}}
\caption{\footnotesize{The 11 single spectra and the
corresponding co-added data (\textit{bottom}) are plotted around the region of
L4R1 and L4P1 (\textit{vertical lines}). The stronger \hhh features can be merely  distinguished
in single spectra.}}
\label{fig:11 spectra 2009}
\end{figure}

The 11 different spectra and the corresponding co-added data (\textit{bottom})
are shown in a region around L4R1 in Figure \ref{fig:11 spectra 2009}.

\subsection{Reduction}
The last version of the UVES pipeline\footnote{Version 4.9.5} has been used  for
 data reduction. The pipeline first uses a set of five  bias
to make a mean bias 
 free of cosmic ray hits which is then
subtracted to all the two dimensional format
 images.
 A pinhole lamp image is used for identifying the precise location of the
Echelle orders which are     curved
and tilted upwards.

The pixel-wavelength conversion of the wavelength calibration  is done by using
 the corresponding  calibration spectrum. 
Murphy et al. (\cite{Murphy08}) and Thompson et al. (\cite{Thompson09a}) 
independently showed  that the
standard Th/Ar line list used in the old UVES pipeline  was a primary
limiting factor. The laboratory wavelengths of the calibration spectrum were
only given to three decimal digits (in units of \AA)
and, in many cases, the wavelengths were truncated rather than
rounded  from four decimal places (see Murphy et al. \cite{Murphy07}).

Thompson et al. (\cite{Thompson09b}) re-calibrated the wavelength solutions using
the 
calibration line spectra taken during the observations of the  QSOs and
argued that the new wavelength calibration was a key element in their null result.

The new data UVES pipeline has solved 
these problems. 

The UVES blue frame comprise 32 orders,  from absolute number
96 to 124,  covering the wavelength range 374-497 nm while the
molecular lines are spread over  18 UVES echelle orders
(from 106-122) covering the wavelength range 380-440 nm. 

 More than  55\% of the  $\approx$ 400 ThAr lines  in
the region were used to calibrate the lamp exposure by means of a polynomial
 of the 5th order. Typical residuals of
the wavelength calibrations were of $\sim$~0.34~\ma \,or $\sim$~24~\ms\,
at
400 nm and  were found symmetrically distributed
around the final wavelength solution at all wavelengths.  By comparison in
Malec et al. (\cite{Malec10}) the wavelength calibration residuals
have been RMS $\sim$ 80 \ms.

In our set of observations,  calibration spectra were taken before
 and after the object spectra for each night. 
Calibration lamps taken  immediately before and after 
object observation provides  accurate monitoring  of  physical  variations. Moreover, the calibration
frames were taken in special mode to avoid automatic spectrograph resetting at the
start of every exposure. Since Dec 2001 UVES
has implemented an automatic resetting of the Cross Disperser
encoder positions at the start of each exposure. This implementation has been
done to have the possibility to use daytime
ThAr calibration frames for saving night time. If this is excellent
for standard observations, it is a problem  for the measurement of fundamental
constants which requires  the best possible wavelength
calibration.
{\chg 
Only calibration spectra that are {\it attached} as template to the OB allow to take the
calibration exposure in exactly the same instrument setting as the
science exposure. These calibrations can be taken upon user's request in addition to the ones
from the calibration plan.}
 Thermal-pressure changes  move in the cross dispersers in different ways,
 thus introducing relative shifts between
the different spectral ranges in  different exposures.

It should be emphasized that this effect has not been taken into account
in  the analysis performed so far on UVES data  for $\mu$ variability.

There are no measurable temperature changes for the short exposures of
the calibration lamps but during the much longer science exposures the
temperature drifts generally by $0.1$ K, and in two cases the drift is of
$0.2$ K while  in other two there is no measurable change. Pressure
values are surveyed at the beginning and end of the exposures
and changes range from 0.2 to 0.8 mbar. The estimates for
UVES are of 50 \ms\,  for $\Delta$T = 0.3 K or a $\Delta$P = 1 mbar
(Kaufer et al. 
\cite{Kaufer2004}), thus assuring a radial velocity
stability within $\sim 50$ \ms.

Individual spectra are corrected for the motion of the observatory
about the barycenter of the Earth-Sun system and then reduced to
vacuum. {\chg The velocity component along the direction to the object of the
barycentric velocity of the observatory was calculated using the
date and time of the midpoint of the integration to minimize the influence of changes.
The changes in radial velocity during exposure induce a symmetric modification of the line profile.
The absorption profile is not strictly Gaussian (or Voight) anymore but rather
slightly squared-shaped 
(since the  FWHM of the line is $\approx$ 5 \kms and the smearing of the line by Earth motions of $\pm$ 40 \ms
the effect is negligible. The line shapes remain symmetric in any case and possible changes of radial velocities
during exposure effects only the quality of the fit, it does not influence the measured centroid of an absorption line.}
The wavelength scale was then corrected
for this motion so that the final wavelengths are vacuum wavelengths as observed
in a reference frame at rest relative to the
barycenter.

 The air wavelengths have been transformed into vacuum by means of the
dispersion formula by Edlen (\cite{Edlen66}). 
Drifts
in the refractive index of air inside the spectrograph between the
ThAr and quasar exposures will therefore cause miscalibrations.
According to
the Edlen formula for the refractive index of air,  temperature and atmospheric pressure changes 
of  1 K and 1 mbar  
would cause differential velocity shifts between 370 nm and 440
nm of $\sim 10$ \ms. 


\section{Preprocessing}
\subsection{Spectral radial velocity shifts}
\begin{figure}[h]
\resizebox{\hsize}{!}{\includegraphics[clip=true]{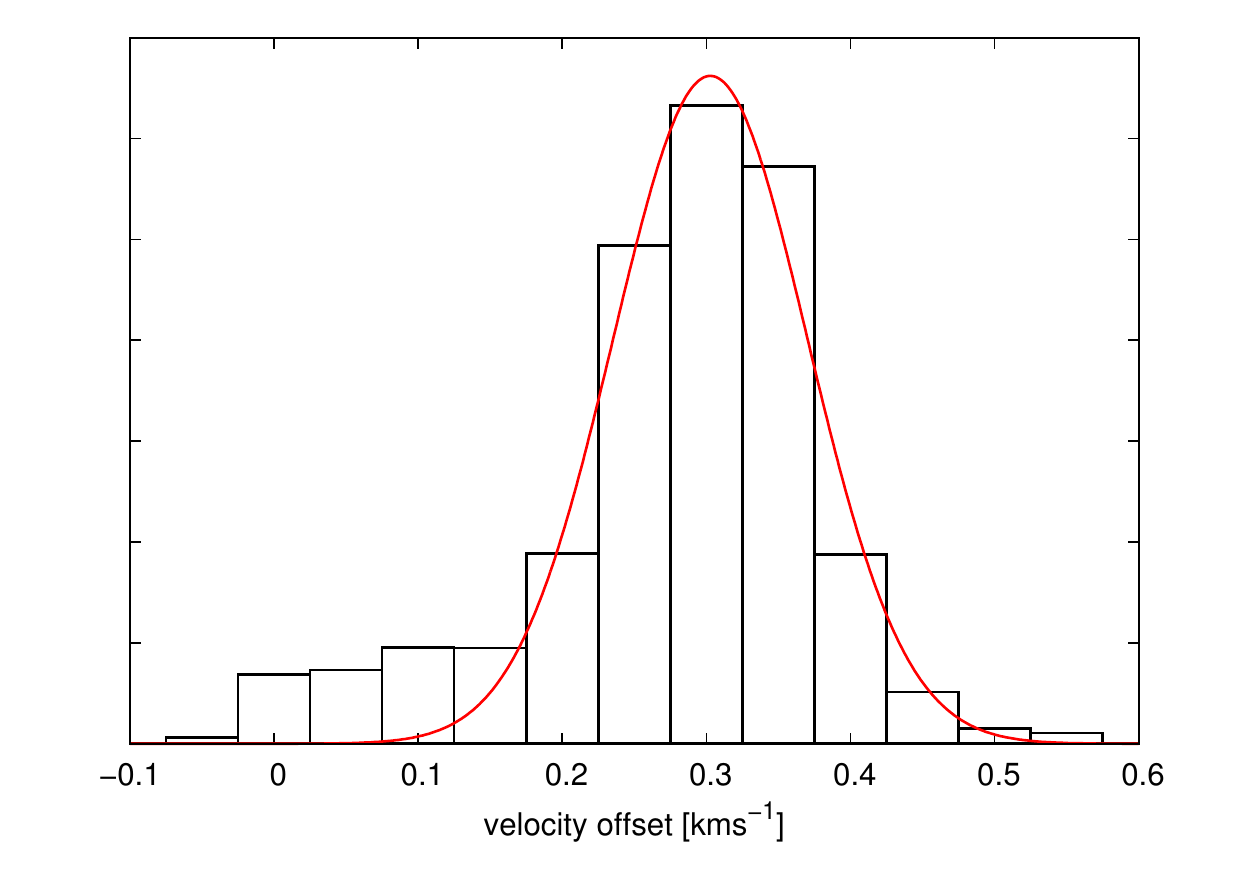}}
\caption{\footnotesize{Exemplary bootstrap histogram of the median position  of
lines
in spectrum 8 with respect to $z = 3.02489817$. The Gaussian fit corresponds to 302 $\pm$ 4 \ms (compare Table \ref{table:1}).}}
\label{fig:shifts_MAD}
\end{figure}

 Wendt and Molaro (2011)  showed the presence of    overall shifts  between  spectra obtained with slit spectrographs such as UVES.   For checking such a possibility in the new data set we obtained a median velocity per each exposure by fitting   as many lines as possible  in each of the   eleven  spectra.
The 50  H$_2$ lines were initially selected for that, but due to the relatively low SN spectra in single exposures, not all could  be fitted, and for instance  most of the weak  J=0 transitions have been  missed.
\begin{table}
\caption{Median velocities of single spectra}
\label{table:1}
\centering          
\begin{tabular}{c c c}     
\hline\hline       
spectrum & median velocity [\kms] & lines considered\\
\\ 
\hline      
1 & 0.075 & 46\\
2 & 0.058 & 42\\
3 & -0.136 & 41\\
4 & 0.223 & 43\\
5 & 0.354 & 47\\
6 & -0.206 & 47\\
7 & -0.015 & 44\\
8 & 0.310 & 48\\
9 & -0.051 & 40\\
10 & -0.107 & 47\\
11 & -0.004 & 48\\
\hline                  
& 0.046&\\
\end{tabular}
\end{table}
The median radial velocity in respect to the chosen absorber redshift derived from all 
detected lines within an individual exposure ranges from - 200 \ms 
to + 354 \ms for the 11 spectra. Due to the low quality
 of the fits to single lines of individual spectra with low signal-to-noise, 
 the median was chosen.
{\chg Despite the low quality of the individual fits 
the median velocities  listed in Table \ref{table:1} are well defined as shown
in Figure \ref{fig:shifts_MAD}.}
Bootstrapping is a useful practice of estimating properties of an estimator (i.e. its error).
For example 48 line positions were determined in spectrum 8 (see Table \ref{table:1}). Their median velocity
offset corresponds to 310 \ms. The bootstrap histograms were implemented by constructing a number of resamples of the observed line positions (and of equal size to the observed data set), each of which is obtained by random sampling with replacement from the original data set.

For the data at hand the obtained velocity offsets bear no statistical
significance because of the large scatter of the individual position measurements per single spectrum.
The resulting offsets are, however, comparably well defined which made us confident   to apply
 this procedure. The described method had no significant impact on the final result at
the current level but it might provide a helpful tool in the future to check for potential
 inter-spectra shifts.
 The resulting
 velocities reflect the offsets of the individual spectra to a 
reference  redshift of the \hhh absorber.
Of the 11 shifts a mean velocity offset (relative to the assumed
z$_{\mathrm{abs}}$)
of 46 \ms was computed\footnote{A non-zero mean radial velocity directly
reflects a deviation from the assumed absorber redshift.}. This low residual offset verifies the assumed
redshift for the absorption system.

\section{Selection of lines and line fitting}
\label{sec:lines}
For the analysis a total set of 50 \hhh lines are
fitted. 
{\chg This preselection is based on earlier line identifications (see Wendt \& Molaro \cite{Wendt11}) including
curve of growth analysis to avoid blends with the Lyman-$\alpha$ forest or other \hhh lines.}
The separate spectra are not coadded in  this step since the fitting algorithm
works on the different data sets simultaneously.
This allows us to omit further rebinning of the data. The fitting code is based on an evolutionary algorithm,
which tracks the global minimum via an interactive process of covering the parameter
space (see Quast et al. \cite{Quast05}). 
Each set of fitting parameters is evaluated via $\chi^2$ in every  single spectra.
Thus, there is no need to redistribute the flux of each exposure to pixels of equal central wavelengths.
Constant velocity offsets, however, potentially influence the shapes of the absorption features as  would
be the case of  coadded spectra. Evolutionary fitting algorithms are less prone to converge in a local minimum
rather than find  the global minimum. An advantage over Monte Carlo chain methods as applied for the purpose
of line fitting in simple cases for example in King et al. (\cite{King08}) 
is the drastically reduced need for computer power. Additionally, the principle of evaluating
 multiple groups of parameters  independently  allows for consequent parallel computing.


For each set of lines sharing the same rotational level a common column density
and a common broadening parameter is fitted with simplified pseudo-Voigt-function profiles.
For weak lines a mere Gaussian profile would suffice since natural line broadening has no noticeable 
impact on the line shapes. In case of QSO 0347-383, only a single component is observed in \hhh.
The only free parameter per each individual line  is the radial velocity with
respect to the absorber redshift.
Out of the 50 \hhh lines analyzed, eight were excluded since they showed a comparably
large positioning error of more than 300 \ms. This is mostly due to a continuum  highly contaminated 
by the presence of  hydrogen absorption in the environment of the affected lines.

%
We note  that our procedure is different from the one followed by King et al. 
(\cite{King08}). They fitted numerous additional components in
a region of \hhh absorption to narrow down the $\chi^2$ of the fit to the data.

\begin{figure}[h]
\resizebox{\hsize}{!}{\includegraphics[clip=true]{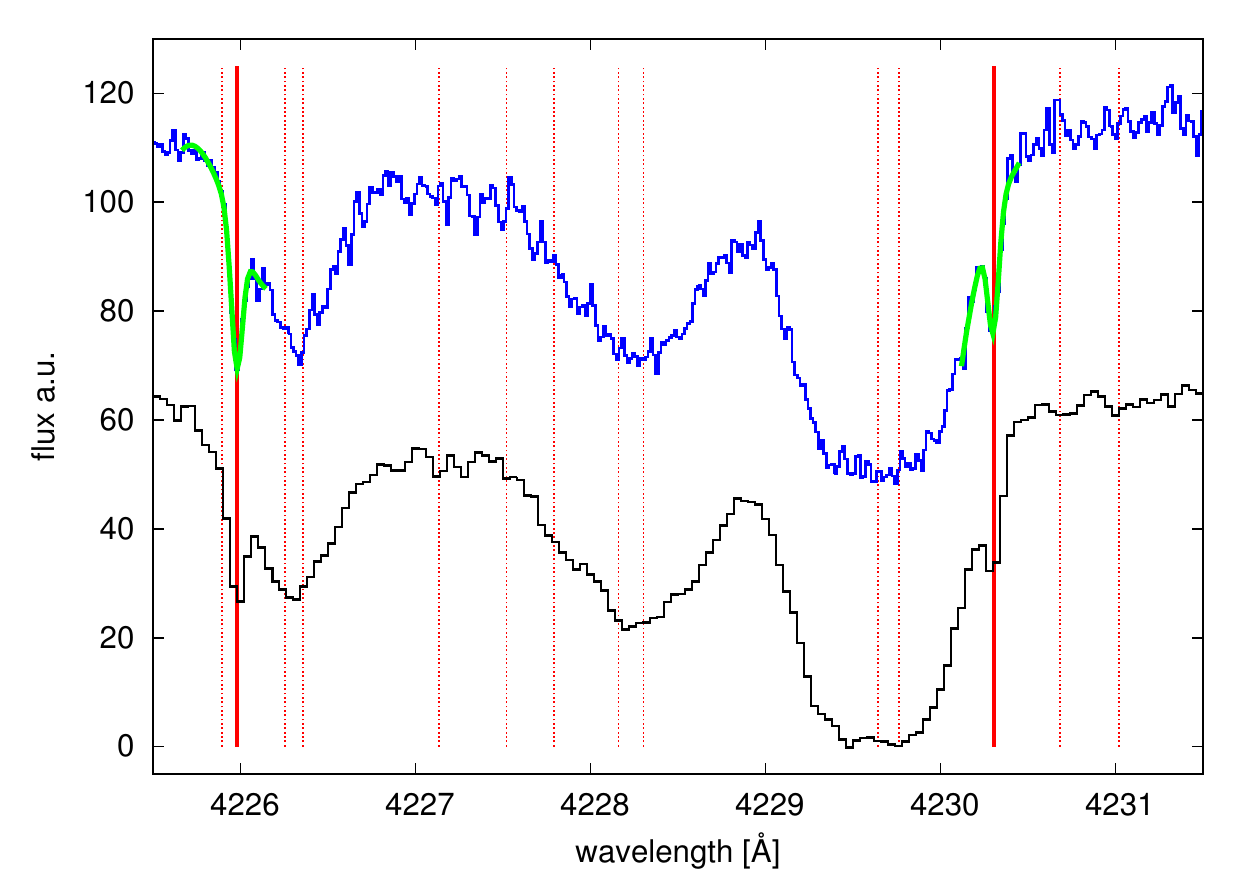}}

\caption{\footnotesize{Comparison between the 2009 data (\textit{top graph})
and, with an offset, the
original single observation run data of 2002 (9 frames, \textit{bottom}) as used
by Ivanchik et al. (\cite{Ivanchik05}), King et al. (\cite{King08}), 
Thompson et al. (\cite{Thompson09a}), and Wendt \& Molaro (\cite{Wendt11}). The vertical lines indicate the
positions of the \hhh component (\textit{solid}) and the 12 additional
lines fitted in King et al. (\cite{King08}). {\chg In green the single line fits (L4R1 and L4P1) with their corresponding local continuum polynomial fit from this work. Note, that the fits for the analysis were not carried out on the rebinned and coadded data as shown in this figure.}}}
\label{fig:2009 king}
\end{figure}

Figure \ref{fig:2009 king} compares a portion of spectrum  used in the analyses by Ivanchik et al. (\cite{Ivanchik05}), King et al. (\cite{King08}), Thompson et al. (\cite{Thompson09a}), and Wendt \& Molaro (\cite{Wendt11}) with the same portion of the new data we are analyzing here. The \textit{solid red} vertical lines mark the \hhh component L4R1 and L4P1
and the \textit{dotted red} lines indicate the 12 additional lines in that
region\footnote{Their individual positions are extracted from the plot in King et al. (\cite{King08}).}. 
The \textit{upper} plot corresponds to the data of 2009 and reveals
that some of the extra components clearly  recreate the flux observed in 2002
but evidently do not correspond to factual properties of the absorber.
%

In   King et al. (\cite{King08}) the evolution of $\chi^2$ with an increasing number of additional free
lines is the main  criterion to fix the total number of components.
While that approach clearly reduces the residuals of the fit, it may not reflect the physical properties
of the absorber. Though it is likely that the absorber structure is too complex to be represented
by a single component,  we  prefer to integrate the uncertainty of the true
nature of the velocity components into the fitting uncertainty rather than 'generating' components to fill
up the flux in a poorly   known continuum. Higher resolution spectra may verify or falsify some of the decisions on additional components and help to distinguish between apparent
precision (lower $\chi^2$) and reached accuracy (better description of the
physical conditions of the absorber) or rather the limit on information on the absorber.

\section{Results}

\begin{figure}[h]
\resizebox{\hsize}{!}{\includegraphics[clip=true]{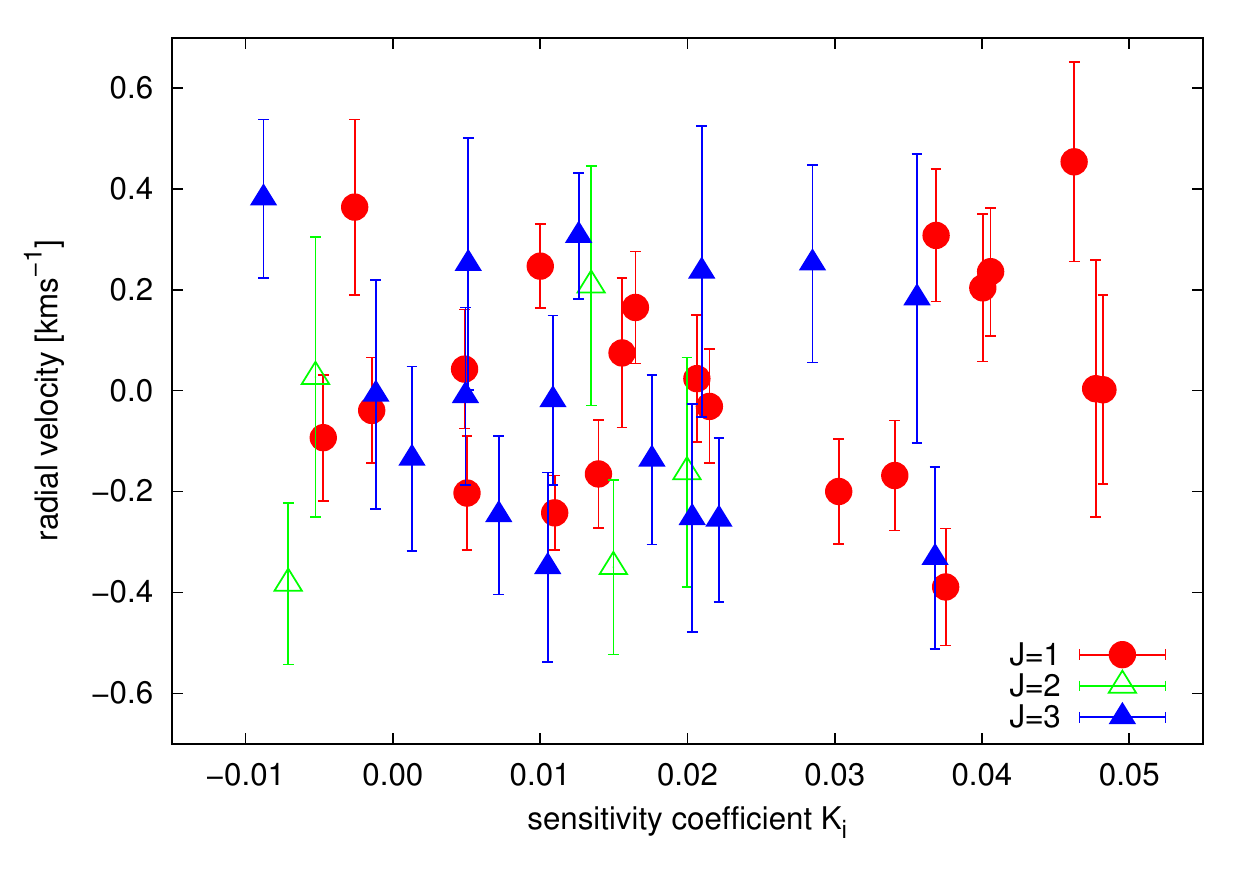}}
\caption{\footnotesize{Measured radial velocity vs. sensitivity for all 42
lines. Any correlation would indicate a change in $\mu$. {\chg The different symbols and colors used correspond
to the observed three rotational levels. The errorbars reflect the 1 $\sigma$ errors.}}}
\label{fig:v_ki}
\end{figure}

In Figure \ref{fig:v_ki} the  measured radial velocities of the 42 \hhh lines are plotted against the sensitivity towards $k$ coefficients  of the corresponding transition. 
Any correlation therein would indicate a variation of $\mu$ 
at z=3.025 with respect to laboratory values. 
{\chg 
Table \ref{tab:lineparamters} lists the broadening parameter and the column density 
for all the lines from one particular J level fit together and consistently with one N and one b
per rotation level.}
\begin{table*}[h]
\caption[]{Line parameters for the observed three rotational levels of \hhh.}
\label{tab:lineparamters}      
\centering          
\begin{tabular}{l l l}
\hline
\hline
J & b [\kms] & logN \\
\hline
1 & 1.34 $\pm$ 0.04 & 14.36 $\pm$ 0.02 \\
2 & 1.39 $\pm$ 0.32 & 13.72 $\pm$ 0.02 \\
3 & 2.02 $\pm$ 0.05 & 13.91 $\pm$ 0.01 \\
\hline
\end{tabular}
\end{table*}
 

%

The data give no hint towards variation of the proton-to-electron mass ratio in the course of cosmic time.
The uncertainties  in the line positions of the \hhh features due to the photon noise are  estimated by the fitting algorithm. These are shown in the errorbars
in Figure \ref{fig:v_ki} and reported in  Table \ref{tab:linelist}.  The mean error  in the line positioning is of 152 \ms. Even at first glance the given errorbars in Figure \ref{fig:v_ki} appear to be too
small to explain the observed scatter.

\begin{figure}[h]
\resizebox{\hsize}{!}{\includegraphics[clip=true]{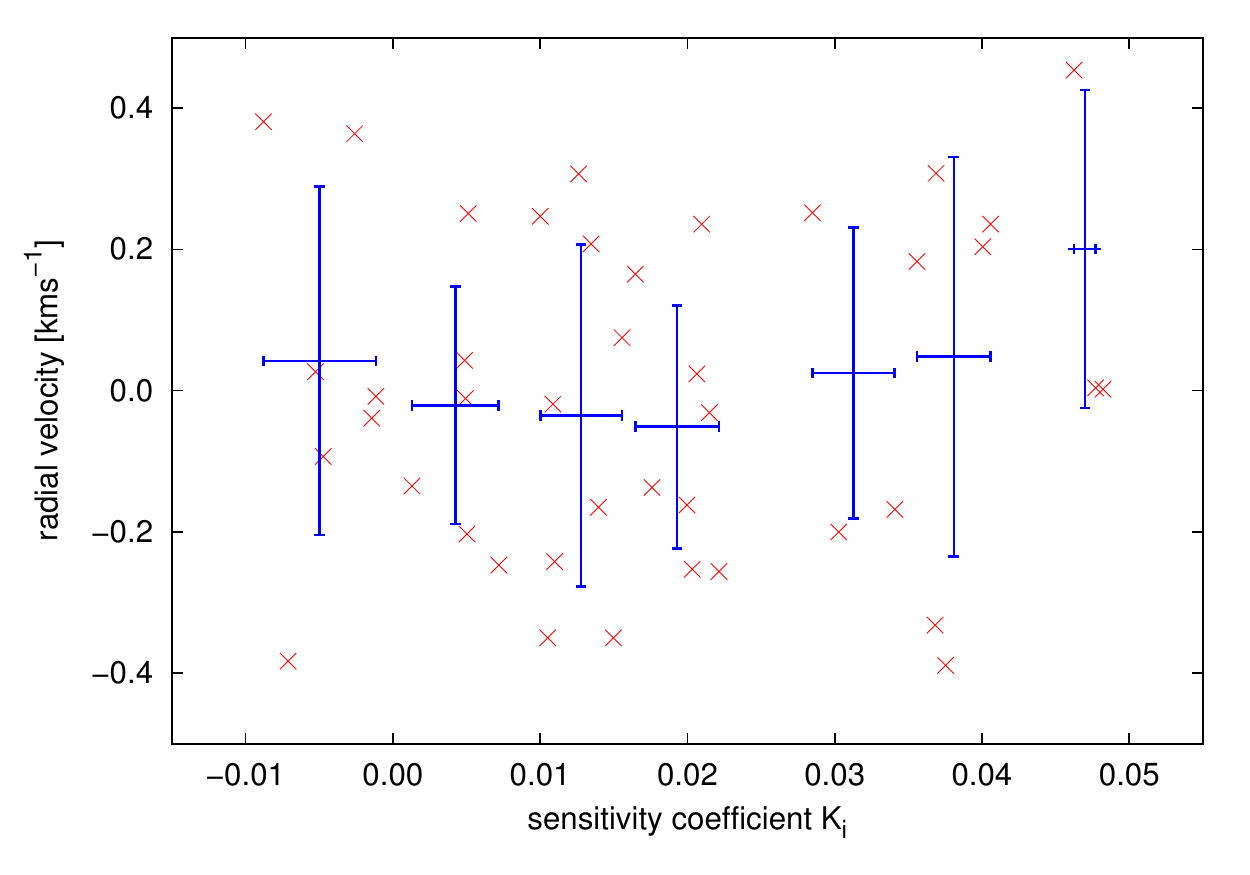}}
\caption{\footnotesize{The observed 42 \hhh lines with their corresponding
radial velocities. In  blue, bins of similar sensitivity are plotted. Blue
y-errorbars reflect the standard deviation of radial velocities within an
interval (x-errorbar).}}
\label{fig:calib_err}
\end{figure}

Figure \ref{fig:calib_err} shows the same line data as is Figure \ref{fig:v_ki} but   lines  with similar $k$ values
are binned.
For a better overview, errorbars are omitted. The red data points (\textit{crosses}) reflect the radial velocity within a small sensitivity range (given as
x-errorbar). The y-errorbars correspond to the standard deviation of the mean
value of the scatter within such a bin.
The scatter within the bins can not be attributed to possible variations of $\mu$
since it is present for basically the same sensitivity parameter. The scatter is
of the order of 220 \ms and thus larger than the positioning error of the individual
lines.
%
That
is also reflected by a reduced $\chi^2$ of 2.7 for a weighted linear fit to the
data  (corresponding to  $\Delta\mu/\mu = (1.8 \pm 8.2) \times
10^{-6}$ at $z_{\mathrm{abs}}=3.025$). The true scatter of
the data is of the order of 220 \ms  and
constitutes an absolute limit of precision. The above mentioned errors of the
fitting procedure require an  additional systematic component to
explain the observed scatter:
\begin{equation}
\sigma_{\mathrm{obs}} \sim \sqrt{\sigma_{\mathrm{pos}}^2 +
\sigma_{\mathrm{sys}}^2},
\end{equation}
with $\sigma_{\mathrm{obs}}$ = 220 \ms, $\sigma_{\mathrm{pos}}$ = 152 \ms, and
$\sigma_{\mathrm{sys}}$ = 151 \ms.

\begin{table*}[h]
\caption[]{List of 42 \hhh lines fitted in QSO 0347-383. Radial velocities given relative to a redshift of $z = 3.02489817$.}
\label{tab:linelist}      
\centering          
\begin{tabular}{c r r r r r r r}
\hline
\hline
Line ID & Lab. wavelength [\AA] & Obs. wavelength [\AA] & Pos. error [\AA] & Velocity [\kms] & Vel. error [\kms]  & $K$ factor\\
\hline
L1P1 & 1094.0519 & 4403.4530 & 0.0026 & 0.364 & 0.174 & -0.003 \\
L1R1 & 1092.7324 & 4398.1360 & 0.0015 & -0.039 & 0.105 & -0.001 \\
L2R1 & 1077.6989 & 4337.6252 & 0.0016 & -0.203 & 0.113 & 0.005 \\
L3P1 & 1064.6053 & 4284.9315 & 0.0012 & 0.247 & 0.083 & 0.010 \\
L3R1 & 1063.4601 & 4280.3151 & 0.0011 & -0.242 & 0.074 & 0.011 \\
L4P1 & 1051.0325 & 4230.2996 & 0.0021 & 0.075 & 0.148 & 0.016 \\
L4R1 & 1049.9597 & 4225.9832 & 0.0016 & 0.165 & 0.111 & 0.016 \\
L5P1 & 1038.1570 & 4178.4767 & 0.0018 & 0.024 & 0.126 & 0.021 \\
L5R1 & 1037.1498 & 4174.4220 & 0.0016 & -0.031 & 0.113 & 0.021 \\
L7R1 & 1013.4369 & 4078.9777 & 0.0014 & -0.200 & 0.104 & 0.030 \\
L8R1 & 1002.4520 & 4034.7650 & 0.0015 & -0.168 & 0.109 & 0.034 \\
L9P1 & 992.8096 & 3995.9617 & 0.0017 & 0.308 & 0.131 & 0.037 \\
L9R1 & 992.0163 & 3992.7595 & 0.0015 & -0.389 & 0.116 & 0.038 \\
L10P1 & 982.8353 & 3955.8147 & 0.0019 & 0.204 & 0.146 & 0.040 \\
L10R1 & 982.0742 & 3952.7519 & 0.0017 & 0.236 & 0.127 & 0.041 \\
L13P1 & 955.7082 & 3846.6281 & 0.0033 & 0.004 & 0.255 & 0.048 \\
L13R1 & 955.0658 & 3844.0424 & 0.0024 & 0.002 & 0.187 & 0.048 \\
L14R1 & 946.9804 & 3811.5054 & 0.0025 & 0.454 & 0.198 & 0.046 \\
W0R1 & 1008.4982 & 4059.1012 & 0.0017 & -0.093 & 0.125 & -0.005 \\
W1Q1 & 986.7980 & 3971.7622 & 0.0016 & 0.043 & 0.118 & 0.005 \\
W2Q1 & 966.0961 & 3888.4364 & 0.0014 & -0.165 & 0.107 & 0.014 \\
\hline
L4P2 & 1053.2842 & 4239.3646 & 0.0034 & 0.208 & 0.237 & 0.013 \\
L4R2 & 1051.4985 & 4232.1695 & 0.0024 & -0.350 & 0.173 & 0.015 \\
L5R2 & 1038.6902 & 4180.6199 & 0.0032 & -0.162 & 0.228 & 0.020 \\
W0Q2 & 1010.9385 & 4068.9194 & 0.0022 & -0.383 & 0.160 & -0.007 \\
W0R2 & 1009.0250 & 4061.2231 & 0.0038 & 0.027 & 0.278 & -0.005 \\
\hline
L2P3 & 1084.5603 & 4365.2445 & 0.0033 & -0.008 & 0.227 & -0.001 \\
L2R3 & 1081.7113 & 4353.7758 & 0.0027 & -0.135 & 0.183 & 0.001 \\
L3P3 & 1070.1408 & 4307.2077 & 0.0025 & -0.011 & 0.176 & 0.005 \\
L3R3 & 1067.4786 & 4296.4891 & 0.0022 & -0.247 & 0.157 & 0.007 \\
L4P3 & 1056.4714 & 4252.1847 & 0.0027 & -0.350 & 0.188 & 0.011 \\
L4R3 & 1053.9761 & 4242.1506 & 0.0018 & 0.307 & 0.125 & 0.013 \\
L5R3 & 1041.1588 & 4190.5564 & 0.0024 & -0.137 & 0.168 & 0.018 \\
L6P3 & 1031.1927 & 4150.4420 & 0.0031 & -0.253 & 0.226 & 0.020 \\
L6R3 & 1028.9866 & 4141.5628 & 0.0023 & -0.256 & 0.163 & 0.022 \\
L8P3 & 1008.3861 & 4058.6547 & 0.0027 & 0.252 & 0.196 & 0.028 \\
L10R3 & 985.9628 & 3968.4022 & 0.0038 & 0.183 & 0.287 & 0.036 \\
L12R3 & 967.6770 & 3894.7970 & 0.0023 & -0.332 & 0.180 & 0.037 \\
W0Q3 & 1012.6796 & 4075.9375 & 0.0021 & 0.381 & 0.157 & -0.009 \\
W1R3 & 987.4487 & 3974.3837 & 0.0033 & 0.251 & 0.250 & 0.005 \\
W2Q3 & 969.0493 & 3900.3246 & 0.0022 & -0.019 & 0.168 & 0.011 \\
W3P3 & 951.6718 & 3830.3853 & 0.0037 & 0.236 & 0.289 & 0.021 \\
\hline
\end{tabular}
\end{table*}
%

%
%
%

A direct linear fit to the unweighted data yields:

\begin{equation}
\Delta\mu/\mu = (4.2 \pm 7.7) \times
10^{-6}.
\end{equation}

\begin{figure}[h]
\resizebox{\hsize}{!}{\includegraphics[clip=true]{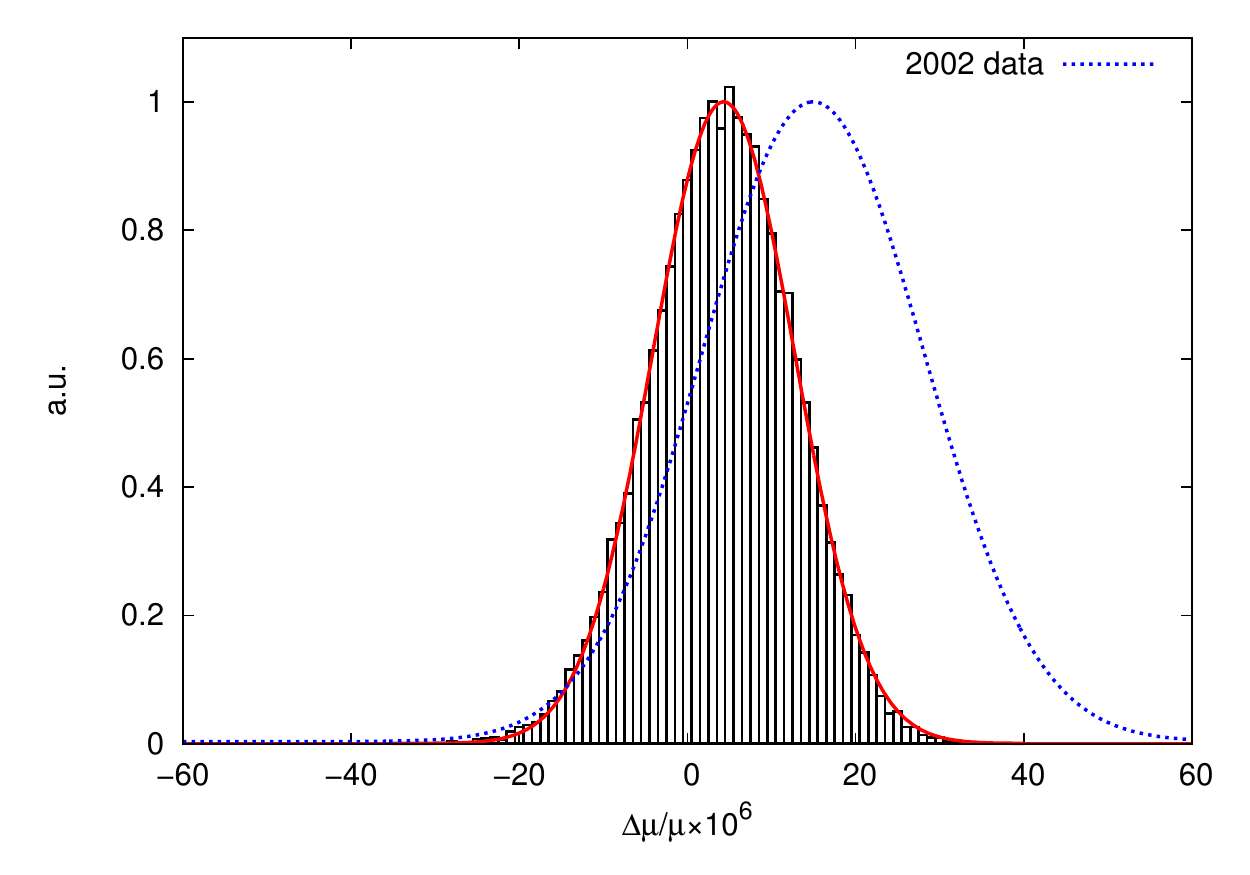}}
\caption{\footnotesize{Bootstrap histogram of 50.000 samples based on all 42 lines.
 The dashed line shows the bootstrap fit of the 2002 data of QSO 0347-383.}}
\label{fig:bootreduced}
\end{figure}

Bootstrap analysis is a robust approach to obtain a linear fit to the data in Figure \ref{fig:v_ki}
 and estimate an error based on the true scatter of the data.
The corresponding bootstrap histogram including Gaussian fit is illustrated
 in Figure \ref{fig:bootreduced}. The Gaussian fit gives:
\begin{equation}
\Delta\mu/\mu = (4.3 \pm 7.2) \times 10^{-6}.
\end{equation}

\section{Systematics}

\begin{figure}[h]
\resizebox{\hsize}{!}{\includegraphics[clip=true]{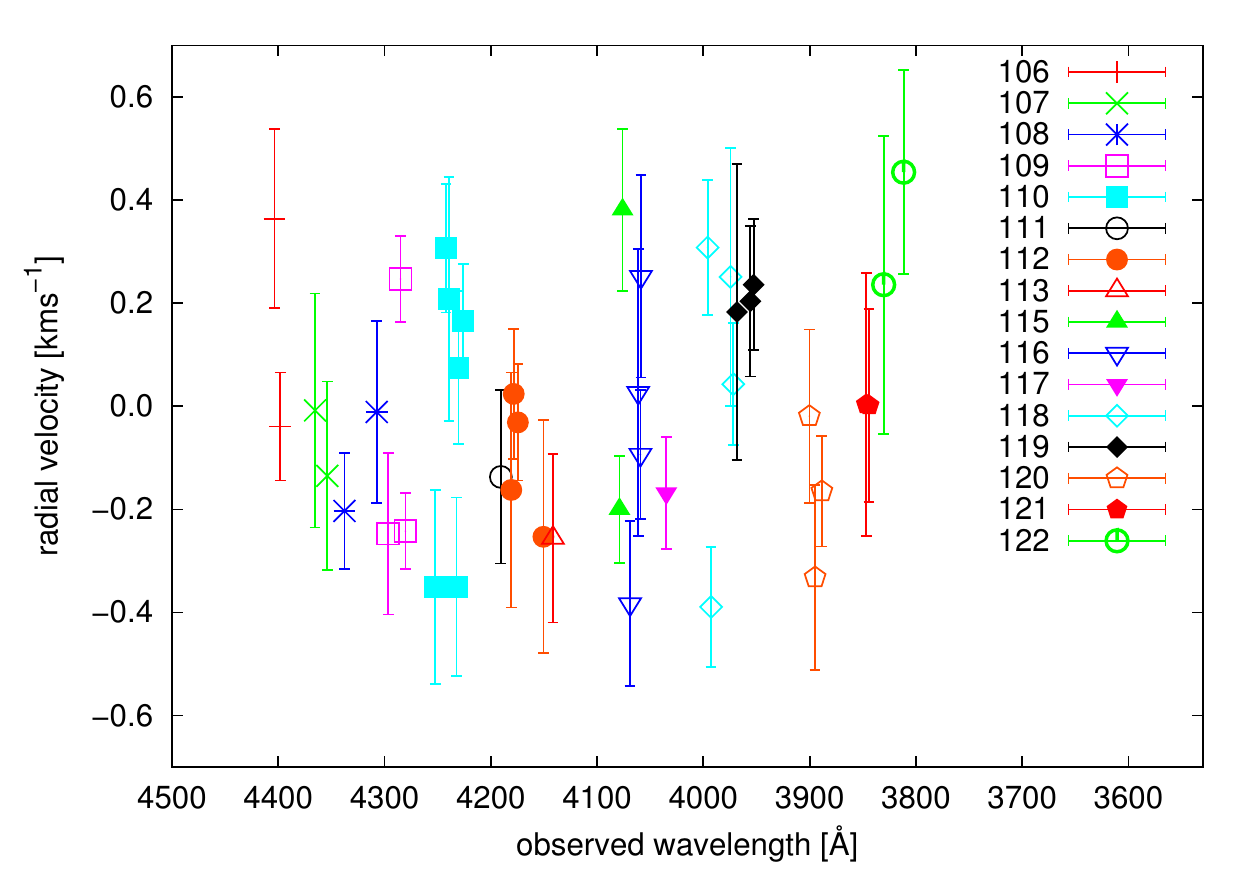}}
  \caption{\footnotesize{All 42 lines with their radial velocity against the
observed wavelength. Different orders are distinguished by  colors and symbols.}}
\label{fig:v_wav_ord}
\end{figure}

Any process which influences the measured
redshift in dependence with the excitation energy would mimic a variation in $\mu$ since the transitions from
higher excited states naturally show a stronger sensitivity towards changes in
$\mu$ (see Varshalovish \& Levshakov \cite{Varshalovich93}).
{\chg Figure \ref{fig:v_wav_ord}  marks the
wavelength ranges covered by the different orders of the CCD spectrum.
Decreasing wavelengths are plotted rightwards since  K sensitivity factors are increasing almost linearly with decreasing wavelengths and therefore Figure \ref{fig:v_wav_ord}  is comparable with Figure \ref{fig:v_ki}.
This new figure  shows no  trend but again a  rather high scatter within the individual
orders can be perceived.}

{\chg In the following, we consider a possible correlation of the line position uncertainty and its
relative position within an order.}
Spectral distortions within the spectral orders have been investigated at the
Keck/HIRES spectrograph by comparing the ThAr wavelength
scale with a second one established from I2-cell observations of a bright
quasar by Griest et al. (\cite{Griest10}). In the wavelength range $\sim 5000-
6200$ \AA\, covered by the iodine cell absorption they found both 
absolute offsets which can be as large as $500 - 1000$ \ms\,  and an 
additional saw-tooth distortion pattern with an amplitude of about $300$ \ms.
The distortions are such that transitions at the order edges appear
at different velocities with respect to transitions at the order
 centers when calibrated with a ThAr exposure. {\chg This would introduce relative velocity shifts between
different absorption features up to a magnitude the analysis with regard to $\Delta\mu/\mu$ is sensitive to.}

 Whitmore et al. (\cite{Whitmore10}) recently repeated the same test for UVES with similar finding
though the saw-tooth distortions show slightly reduced peak-to-peak velocity
variations of $\sim 200$ \ms. The physical explanation for those distortions
is not yet known, so it is still to be examined  whether the deviations are the
same at other wavelengths or depend on the specific exposure.

The available solar atlas can be used to check UVES interorder distorsions as suggested in   Molaro \& Centurion (\cite{Molaro11}).
 For this purpose 
UVES observations of the solar spectrum reflected by the  asteroid Iris   were  taken on  Sep 2009  with a  resolving power  R $\approx$ = 85000. 
 
\begin{figure}[h]
\resizebox{\hsize}{!}{\includegraphics[clip=true]{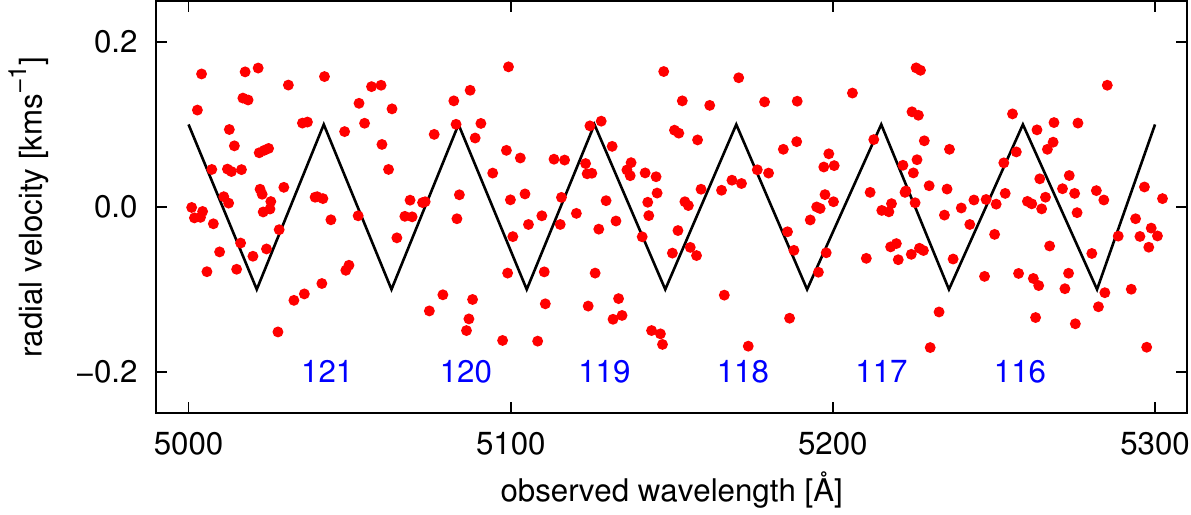}}
\caption{\footnotesize{Radial velocity shifts of individual lines Molaro et al. (2011).}}
\label{fig:fig4}
\end{figure}

The differences  of the positions in the UVES  spectrum   and  the absolute positions of the  solar atlas  for 238
 solar photospheric lines in the region between 500-530 nm  (Orders 121 to 116) are  shown in Fig  \ref{fig:fig4}.
   The schematic saw-tooth pattern detected  by Whitmore et al. (2010)   is also sketched in the figure.
 {\chg The stochastic distribution of the data points does not allow for any conclusion to be drawn about an underlying  pattern.}
%
%
The observed dispersion  is of 82 \ms. Since the typical error in the   measurement  of  lines  in the UVES Iris spectrum is of $\approx$  30  \ms and wavelength calibration residuals are of 25 \ms,  there is an excess in the observed  dispersion which suggests  the presence of local deviations in the UVES spectrum.
 The   saw-tooth pattern detected  by Whitmore et al. (2010)   is   not revealed  by our test.
 
%
%
%

\begin{figure}[h]
\resizebox{\hsize}{!}{\includegraphics[clip=true]{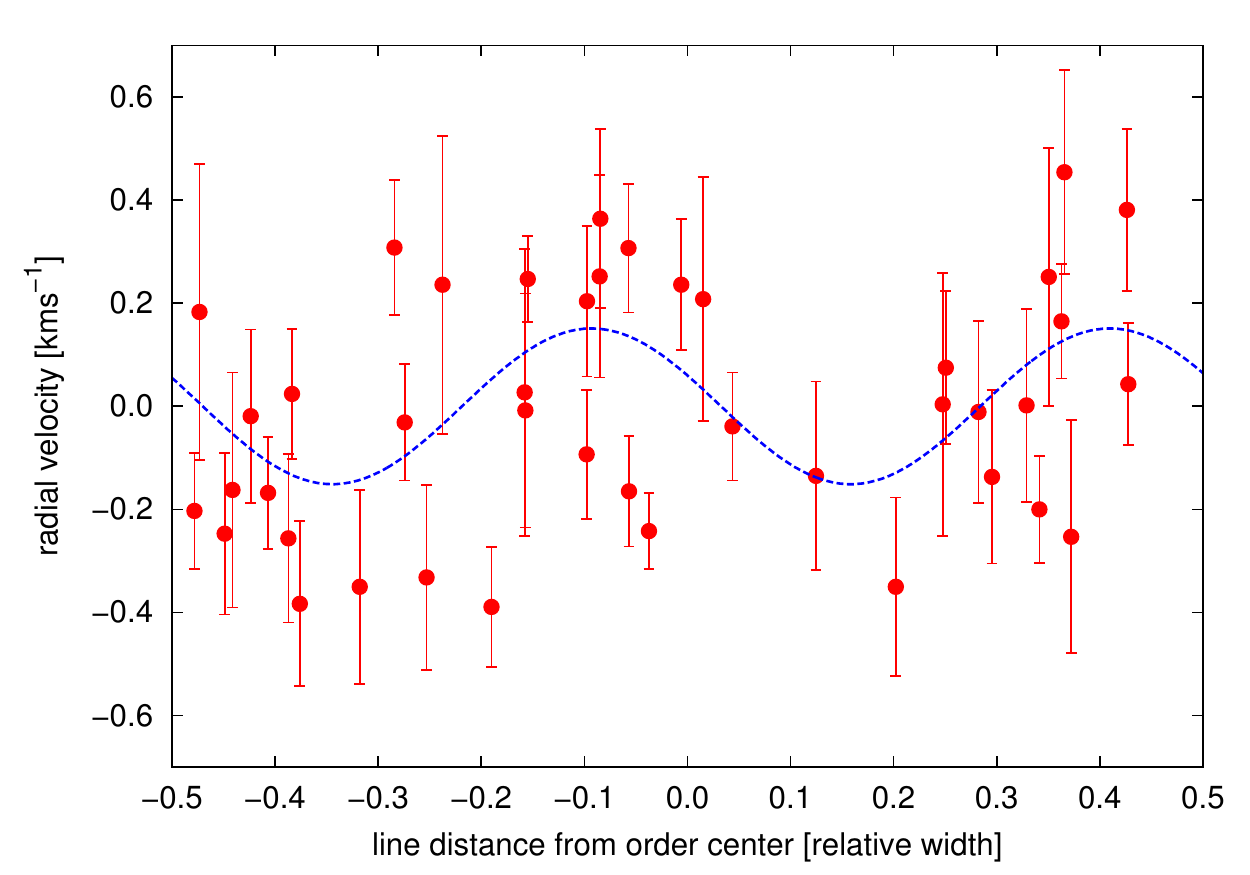}}
\caption{\footnotesize{All 42 lines with their radial velocity against their
relative position within their order. A cosine fit with an amplitude of 151 \ms
is shown in blue.}}
\label{fig:v_dwav}
\end{figure}

Figure
\ref{fig:v_dwav} sorts the 
observed lines according to their relative position within their order. The
origin of the 
abscissa reflects the central position within an order. 
All observed 
orders are stretched to an identical scale and over-plotted for this purpose.
Lines near -1/2 on the
X-axis are positioned near the left rim of the order, and so on.
The distribution of obtained radial velocities seems to show a certain periodic
pattern. The blue curve  shows a fitted cosine with an amplitude of 151 \ms. Considering the errorbars of the individual lines, this is no more than a slight indication which supports  the presence of local distortions resulting from a non perfect calibration.
%
Systematic errors at the level of
few hundred \ms\,  have been revealed also in the UVES data by
comparison of relative shifts of lines with comparable response
to changes of fundamental constants (Centurion et al. \cite{Centurion09}) and Molaro et al (2012).
Molaro et al (2012) suggest that these distortions  may originate from the block stiching of the CCDs. 
{\chg A  CCD device is built-up
by means of several sub-unit  blocks with typical sizes of 512 pixels. The stitching of the blocks process produces
misalignments of the order of few 0.01 of the pixel size in the block conjunctions.  The ThAr has not enough lines to follow these imperfections  which are therefore flattened  in the pixel-to-wavelength conversion by a low order polynomial resulting into the observed  spectral distortions.}%
\section{Discussion}
The  result of $\Delta\mu/\mu = (4.3 \pm 7.2) \times 10^{-6}$ we obtained is consistent with no variation of $\mu$ between
 $z_{\mathrm{abs}}=3.025$ and $z=0$. The null-result is in agreement with recent
 publications on $\Delta\mu/\mu$ by King et al. (\cite{King11}) and Weerdenburg et al. (\cite{Weerdenburg11}) at $z_{\mathrm{abs}}=2.811$ and $z_{\mathrm{abs}}=2.059$, respectively.
However, the present work utilizes the line-by-line fitting method (as i.e. in Ivanchik et al. \cite{Ivanchik05})
 in contrast to the other works which applied a 
comprehensive fitting method (CFM). The \hhh system 
in the spectrum of QSO 0347-383 has the particular advantage of comprising a single velocity component,
which renders observed transitions independent of each other. For absorption systems with two or more closely and not properly resolved velocity components  many systematic errors may influence 
distinct wavelength areas. The CFM  fits all \hhh components along with additional H$_I$ lines and handles an artificially applied $\Delta\mu/\mu$ as free parameter in the fit. 
{\chg The best matching  $\Delta\mu/\mu$ is then derived via the resulting $\chi^2$ curve. The CFM aims to achieve the lowest possible $\chi^2_\nu$ via additional velocity components.
 In this approach the information of individual transitions is lost though since merely the overall quality of the 
comprehensive model is judged. 

In Weerdenburg et al. (\cite{Weerdenburg11}) the number of velocity components is increased as long as 
the composite residuals of several selected absorption lines differ from flat noise.
The residuals therein do not take into account the known inaccuracy of the estimated flux error
 (see, i.e., Wendt \& Molaro \cite{Wendt11}, King et al. \cite{King11}). }

As pointed out by King et al. (\cite{King11}), for multi-component structures with overlapping velocity
components the errors  in the line centroids are heavily correlated and a simple $\chi^2$ regression
is no longer valid. The same principle applies for co-added spectra with relative velocity shifts.
The required rebinning of the contributing data sets implements further autocorrelation of the individual
'pixels'. 

The uncertainties of the oscillator strengths $f_i$ that are stated to be up to 50\% (Weerdenburg et al. \cite{Weerdenburg11})  might further affect the criteria for additional velocity components.
The method of CFM was applied for QSO 0347-383 by King et al. (\cite{King08}). Section \ref{sec:lines}  and in particular Figure \ref{fig:2009 king} reveal some of the mentioned difficulties. 
{\chg The approach to fit individual lines with common physical properties which was applied here
 allows us to carry out an error analysis which reproduces the impact of differential shifts within
 the spectral orders. This yields a higher transparency of the error-budget
 for individual lines but at the possible cost of larger scatter resulting in a slightly larger error-estimate. 
For the small number of \hhh lines observed in the spectrum of QSO 0347-383 we prefer the method applied in this work.}
The immediate advantage is that we are not forced to estimate the different systematic errors based on assumptions. Instead the true limiting error can be gathered directly from the data distribution
and we are further able to attribute it to different sources.
%

{\chg The new set of UVES observations of QSO 0347-383 this analysis is based has been taken with special
 care aimed to improve the measurement of  $\Delta\mu/\mu$ in a robust manner. }
In particular the observations have been taken with higher resolution, a $1 \times 1$ binning
and  calibration lamp spectra in direct combination with the main exposures. These  boost the precision
of the analysis roughly by a factor  two  with comparison  with Wendt et al. (\cite{Wendt08}).
 We have shown that at the  current level, calibration issues  become the dominant source of error. In addition to positioning errors,
 which are  related to   the signal-to-noise-ratio of the data, we observe for the first time inter order
distortions which seem to be of the same order of magnitude of the
 uncertainties in the line positions. The  conclusion is that we do not detect change in
the value of $\mu$ to 1 part in 10$^5$ over a time span of 11.5 Gyr, which is  approximately 80\% of the age of the universe. 

{\chg High resolution data and attached calibration spectra are the key to understanding and handling the systematics
which limit the precision of $\Delta\mu/\mu$ measurements. It is important to fully control the 
 analysis of individual absorption systems and to minimize the errors involved wherever possible before 
extending the $\Delta\mu/\mu$ analysis to multiple systems. Different characteristics of  individual  absorbers
 tend to get lost while not all errors of the measurements are likely to average out.

Until new ways of wavelength calibration such as optical laser frequency combs (see i.e. Steinmetz et al. \cite{Steinmetz08}) are installed for large optics, the data at hand
is of the best quality available. High resolution VLT-data with special care regarding
the calibration frames allows for the best precision that can be reached nowadays.
In the  context  of $\Delta\mu/\mu$ measurements, optical spectra of \hhh at high redshifts are still
 of high importance to complement high precision determinations of $\Delta\mu/\mu$ in the local universe
via observations in the radio regime.}
\begin{acknowledgements}
We are thankful for helpful discussions on this topic with S.A. Levshakov and D. Reimers.
\end{acknowledgements}

\bibliographystyle{aa}

\end{document}